\documentclass[aps,pra,reprint]{revtex4-1}
\usepackage{natbib}
\usepackage{graphicx}
\usepackage{epstopdf}
\usepackage[caption=false]{subfig}
\usepackage{bm}
\usepackage{amsmath}
\usepackage{hyperref}

\graphicspath{ {./Figures/} }

\newcommand{\h}{\mathcal{H}}
\renewcommand{\vec}[1]{\mathbf{#1}}  

\begin{document}
\title{A Simple Bayesian Method for Improved Analysis of Quasi-2D Scattering Data}
\author{Alexander T Holmes}
\email{a.t.holmes@bham.ac.uk}
\affiliation{School of Physics and Astronomy, University of Birmingham, Birmingham B15 2TT, UK}

\date{\today}
\begin{abstract}
A new method is presented for the analysis of small angle neutron scattering data from quasi-2D systems such as flux lattices, Skyrmion lattices, and aligned liquid crystals. A significant increase in signal to noise ratio, and a natural application of the Lorentz factor can be achieved by taking advantage of the knowledge that all relevant scattering is centered on a plane in reciprocal space.  The Bayesian form ensures that missing information is treated in a controlled way and can be subsequently included in the analysis. A simple algorithm based on Gaussian probability assumptions is provided which provides very satisfactory results.  Finally, it is argued that a generalised model-independent Bayesian data analysis method would be highly advantageous for the processing of neutron and x-ray scattering data.
\end{abstract}

\maketitle
\section{Introduction}
The study of flux line lattices (FLLs) in type-II superconductors is a major use of the small angle neutron scattering (SANS) technique \cite{Dewhurst06,Eskildsen11a,Eskildsen11b}. In their so-called `mixed state' these materials contain a periodic distribution of magnetic field. Due to their wave nature and instrinsic magnetic moment, neutrons diffract from these magnetic structures, in a manner which make them a powerful tool for the investigation of FLLs.

The mixed state occurs when type-II superconductors are subject to an external magnetic field $H$ above their lower critical field $H_{c1}$. In this phase, the field penetrates the bulk superconductor, creating regions of suppressed order parameter, surrounded by screening supercurrents. Coherence of the superconducting pair wave-function means that each of these regions, known as vortices or flux lines, corresponds to a single flux quantum $\Phi_0=h/2e$. Repulsion between flux lines causes a 2D lattice to form, which would be close-packed hexagonal in a perfectly isotropic system \cite{Kleiner64}. In real materials, however, the nature of the superconducting pairing and anisotropies in the Fermi surface can cause distortions and even phase transitions in the flux lattice. For example in single crystal niobium, temperature and field-dependent transitions are found in the FLL \cite{Laver06}, which even reflect topological considerations due to the so-called `hairy-ball theorem' \cite{Laver10}.

SANS can give information about the structure of the FLL, and also about the field contrast and lattice perfection.  The contrast is related to the London penetration depth, and hence superfluid density. The way this varies as the temperature approaches zero can tell us about the presence of nodes in the superconducting gap, indicating unconventional pairing symmetry. The lattice perfection is affected by pinning of the flux lines and by thermal fluctuations.

In a uniform field in most materials the FLL is essentially 2-D in nature, consisting of parallel rods aligned with the magnetic field. The 2-D array of rods gives a single plane of spots in reciprocal space, at positions $\vec{q}$ equal to the reciprocal lattice points $\vec{G}_{hk}$.  The width of the spots perpendicular to the reciprocal lattice plane is given by the correlation length along the field direction in real space, and their integrated intensity is determined by the Fourier component of the field contrast $B(\vec{q})$.

\begin{figure*}
  \subfloat[]{\includegraphics[width=0.67\columnwidth]{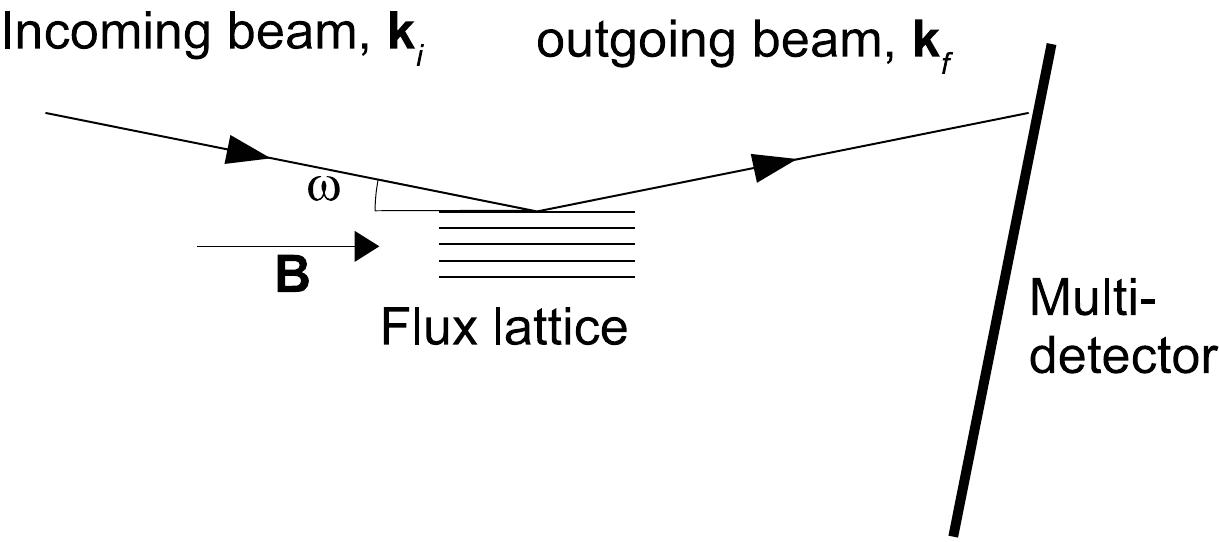}\label{fig:geometry}}
  \subfloat[]{\includegraphics[width=0.67\columnwidth,clip=true,trim=0 3.5cm 0 0]{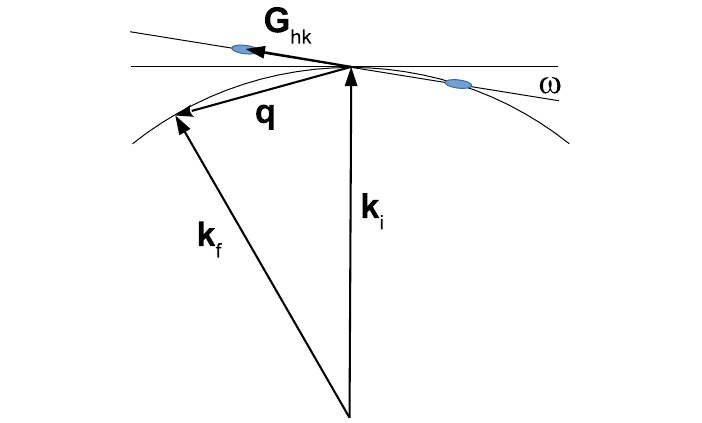}\label{fig:crosssection}}
  \subfloat[]{\includegraphics[width=0.67\columnwidth]{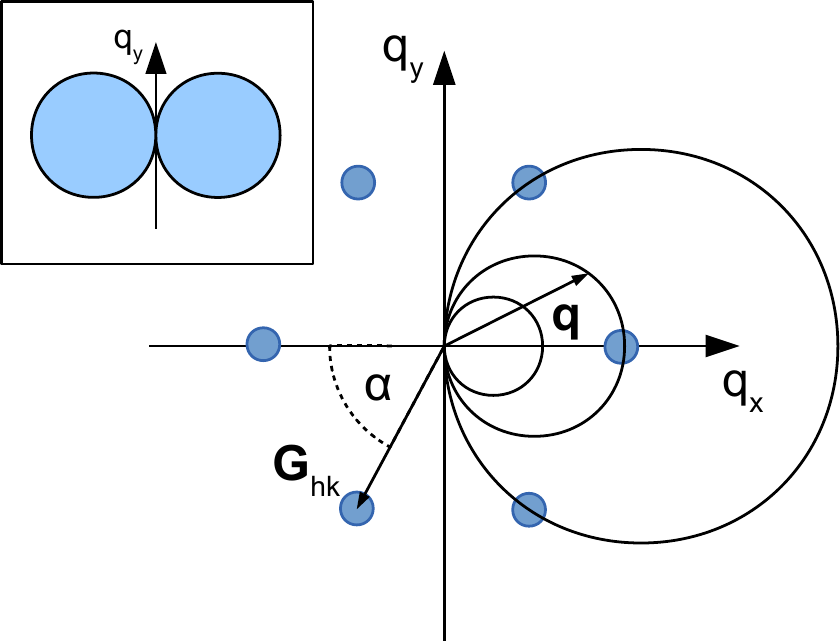} \label{fig:plane}}
  \caption{(Color online) \protect\subref{fig:geometry} Schematic diagram of a typical experimental setup. The incoming neutrons have a wavevector $\vec{k}_i$, outgoing neutrons have wavevector $\vec{k_f}$, where $\left|\vec{k}_f\right|=\left|\vec{k}_i\right|=2\pi/\lambda$, and $\lambda$ is the neutron de Broglie wavelength. Each pixel in the multi-detector corresponds to a different direction of $\vec{k}_f$. \protect\subref{fig:crosssection} Section through the Ewald sphere and reciprocal lattice in the scattering plane. The field and hence reciprocal lattice is rotated by an angle $\omega$ with respect to $\vec{k}_i$. \protect\subref{fig:plane} Reciprocal lattice plane showing circles where the Ewald sphere intersects for different rotations $\omega$ around $\vec{q}_y$. If all scattered intensity is in this plane, then only pixels close to a circle will contain relevant data. The inset in \protect\subref{fig:plane} shows the region of reciprocal space for which relevant data is collected when the field is `rocked' from $-\omega$ to $+\omega$ about single axis in the $y$-direction.}\label{fig:Setup}
\end{figure*}

Figure \ref{fig:Setup}\subref{fig:geometry} shows the experimental geometry; Figs.~\ref{fig:Setup}\subref{fig:crosssection} and \ref{fig:Setup}\subref{fig:plane} show the reciprocal lattice and Ewald sphere relevant to the FLL problem. Diffraction will only occur when the momentum transfer between the neutron and lattice $\hbar\vec{q} = \hbar\left(\vec{k}_f-\vec{k}_i\right)$ is equal to a point in the reciprocal lattice $\vec{G}_{hk}$, where $\vec{k}_i$ and $\vec{k}_f$ are respectively the incoming and outgoing neutron wavevectors. For elastic scattering (where $\left|\vec{k}_f\right|=\left|\vec{k}_i\right|$) the Ewald sphere construction is often used to illustrate this, and in the case of the 2D FLL, it implies that the only diffraction signal to occur will be on the locus of points where the sphere determined by $\left|\vec{k}_i\right|$ intersects the reciprocal lattice plane containing $\vec{G}_{hk}$. This is a circle passing through the origin at $\vec{q}=0$, with a radius depending on $\left|\vec{k}_i\right|$ and the angle between the lattice plane and beam direction. The points where this circle passes through a reciprocal lattice point $\vec{G}_{hk}$ correspond to the Bragg condition for a given lattice point. However the present method makes no assumptions about the form of the lattice, and works for any form of 2D scattering, for example from an aligned nematic liquid crystal \cite{vandenPol09}. It would also work in non-2D cases where other points in reciprocal space are far enough away not to cut the Ewald sphere.

The effect of the finite correlation length along the field direction is to thicken the lattice plane, causing the spots to have a finite width in the direction normal to the plane. The wavelength spread, beam size and divergence also serve to broaden the final diffracted beam, as does mosaic spread of the sample \cite{Cubitt92}.  If the correlation length is the dominant source of broadening then the spreading will have a Lorentzian form in reciprocal space, however if instrumental resolution is the main factor, it is likely to be more Gaussian. For this example we will assume a Lorentzian shape with full width half maximum (FWHM) $\Gamma$, but a different function is straightforward to implement.

During a diffraction experiment, the angle between the incoming neutron beam and the magnetic field is varied by rotating the entire sample environment, including sample and magnet, about an axis passing through the sample. This has the effect of bringing different reciprocal lattice spots in and out of the Bragg condition, as the Ewald sphere passes through them. In general the field and sample can be rotated about two perpendicular axes, by angles referred to as $\omega$ and $\phi$, to access different regions of reciprocal space. We will mostly refer to rotation about a single axis for clarity, but this method works for any combination of $\omega$ and $\phi$. Strictly speaking we should specify an order in which the rotations occur, as they do not commute, however if the angles are small this will not be important. A 2D detector placed several metres away records the diffracted intensity at small angles ($\sim1^\circ$) to the beam axis.  To produce an entire diffraction pattern, a series of measurements are taken at different rotation angles.

\section{Data analysis}
The traditional method of producing a diffraction pattern from such an experiment is to sum the counts on each pixel in the multi-detector over frames (i.e. individual exposures at given rotations) taken over a range of angles which encompass the Bragg condition for all relevant diffraction spots.  Backgrounds are subtracted using measurements taken either in zero field, or above the superconducting transition temperature. This is necessary as the diffracted intensity from the FLL is relatively weak, and superimposed on a $|q|^{-4}$ background. The $\vec{q}=0$ point is determined by a measurement of the undiffracted `direct' beam, which is also used to determine the absolute diffracted intensity in subsequent processing. The counts are also usually normalized to reflect the integration time, or total neutron flux, giving a result in `counts per standard monitor' which accounts for different measuring times.

While simple to understand and implement, the simple `rocking sum' described has a number of disadvantages.  One is that the rate at which the Ewald sphere will pass through a spot depends on the angle between the relevant reciprocal lattice vector and the rotation axis.  As shown in figure \ref{fig:Setup}\subref{fig:crosssection} and \subref{fig:plane}, the reciprocal lattice plane is initially tangent to the Ewald sphere at $\omega = 0$. Then the plane is rotated by $\omega$ about an axis, also within the plane, passing through the point where they touched.  The intersection between the sphere and plane is a circle, fixed at one point on its circumference, expanding in radius in a direction perpendicular to the axis of rotation. The circle will take the smallest range of $\omega$ to pass through a finite sized spot in this direction and will cross the other spots over an angular range larger by a factor of $1/|\cos(\alpha)|$, where $\alpha$ is the angle between the $\vec{q}$-vector of the spot and the perpendicular to the rotation axis.  This is known as the Lorentz factor, and means that spots closer to the rotation axis will have have a correspondingly higher integrated intensity.  One must therefore divide this by the Lorentz factor to calculate the total integrated intensity for individual spots in $\vec{q}$ space.

The other problem is that an unweighted sum of detector counts will include many angles for which any individual pixel is not at or close to the Bragg condition. In that case noise will be accumulated, but no signal. This is an especially big problem when the quality of the lattice is very good, with a long correlation length, as each spot will only appear in a narrow range of angles (with a FWHM known as its `rocking curve width').

One way of handling the noise issue described above would be to do a weighted sum of counts, instead of equally weighting the counts at each rocking angle. The intensity in a given spot will have a Lorentzian shape as a function of rocking angle, with a peak centered at the relevant Bragg angle, if the intrinsic longitudinal broadening of the flux lattice diffraction pattern is dominated by the longitudinal coherence length. The weights for each rocking angle can be optimized so as to reflect the fraction of the peak intensity captured at each pixel for a given frame, if the rocking curve width has a known value.

This method is effectively a fit to the peak intensity. However it runs into problems when the Bragg condition is never reached.  This will happen close to the rocking axis, where the Ewald sphere never cuts the reciprocal lattice plane.  In that case, every single weighting coefficient will be close to zero, and subsequent normalisation will only serve to amplify the noise in these areas.

There are several ways this can be handled, but one of the most justifiable and logically consistent is to use Bayes' theorem, which brings with it several advantages, as we will discuss below.

\section{Bayesian inference of diffraction intensity}
The problem can be stated in the following terms.  We wish to infer the integrated diffracted intensity $I(\vec{q})$ for a region of reciprocal space at a given momentum transfer $\vec{q}=(q_x, q_y, 0)$ in the reciprocal lattice plane (i.e. the plane in reciprocal space passing through (0, 0, 0) perpendicular to the field direction, which is defined as the (0, 0, 1) direction). This coordinate system is fixed with respect to the field and sample; the rocking angle $\omega$ defines the beam direction $\vec{k}_i$, and the detector pixel position determines $\vec{k}_f$ and hence the momentum transfer $\vec{q}=(q_x, q_y, q_z)$. In the simplest case we will make one assumption: that the diffraction pattern is two dimensional, except for a constant longitudinal correlation length, over the entire $q$ range, i.e. that all diffracted intensity can be described by the in-plane intensity and a single Lorentzian or Gaussian maximum perpendicular to the plane, such that
\begin{equation}\label{2d}
I_{3D}(q_x,q_y,q_z)=f(q_z)I_{2D}(q_x,q_y,0),
\end{equation}
where a Lorentzian rocking curve shape gives
\begin{equation}\label{lorentzian}
f(q_z)=\frac{1}{\pi}\frac{\frac{1}{2}\Gamma}{(\frac{1}{2}\Gamma)^2+q_z^2},
\end{equation}
or a Gaussian gives
\begin{equation}\label{gaussian}
f(q_z)=\frac{1}{\sqrt{2\pi}\gamma}\exp\left(\frac{q_z^2}{2\gamma^2}\right),
\end{equation}
with $\Gamma$ being the FWHM in Eq.~\eqref{lorentzian} and $\gamma$ being the r.m.s.\ width for Eq.~\eqref{gaussian}. We will refer to the former from now on, but all considerations apply equally to a Gaussian shape. $f(q_z)$ is normalized so that $\int^\infty_\infty f(q_z) dq_z=1$. It should be noted that $f(q_z)$ has units of $[1/q_z]$.

This model makes no assumptions about the form of the in-plane intensity, either the the diffraction pattern itself, or resolution effects.  It explicitly rules out any modulation other than a correlation length $2/\Gamma$ along the $z$ direction which is the same for all Bragg planes of the FLL, though it could easily be extended to account for anisotropic $\Gamma$.

We have a set of data consisting of an array of counts from a 2D detector taken at a set of rocking angles $\{\omega_j\}$, already normalized and background subtracted (n.b. sets of multiple values or data points are denoted by curly brackets). The intensity at pixel $i$ in frame $j$ will be referred to as $D_{ij}$, with an error $\sigma_{ij}$ determined from Poisson statistics.

In terms of an inference problem, we wish to know the probability, for a given set of points at $\{\vec{q}_i\}$, that the integrated intensity has a particular value $\{I_i\}$, given the experimental data $\{D_{ij}\}$, and a FWHM $\Gamma$.  For convenience and clarity we will select $\{\vec{q}_i\}$ to be equivalent to the detector pixels, which for a given $\left|k\right|$ and small rotation angles can be assumed to have the same $q_x,q_y$ for each $\omega_i$.  We will consider the case in which the neutrons are selected to be (approximately) monochromatic by a velocity selector. Time of flight mode, when a large range of neutron wavelengths is present during a measurement, can also be treated in the same way, but in that case the choice of $\vec{q}$ values at which to calculate the intensity is not as straightforward.

The expected measurement at pixel $i$ in frame $j$ is given by $f_{ij}I_i$ according to Eq.~\eqref{2d}.  This is a function of $(q_x,q_y)_i$, the integrated intensity $I_i$, rocking angle $\omega_j$ and rocking curve width $\Gamma$ in reciprocal space. $\Gamma$ can initially be estimated from the angular rocking curve width $\eta$ for a particular flux lattice diffraction spot (later we will make this a parameter to be determined).

The FWHM $\Gamma$ along the field direction in \AA$^{-1}$ for a spot centred on $\vec{q}=(q_x,q_y,0)$ and rocking axis parallel to the vector $\vec{r}$ is given by
\begin{equation}\label{rockwidthtoq}
    \Gamma = 2 \frac{\left|\vec{q}\times\vec{r}\right|} {\left|\vec{r}\right|} \tan(\eta / 2).
\end{equation}
Where $\eta$ is the rocking curve FWHM in degrees.  Conversely $\eta$ is given by:
\begin{equation}\label{qtorockwidth}
    \eta= 2\tan^{-1}\left(\frac{\Gamma \left|\vec{r}\right|}{2\left|\vec{q}\times\vec{r}\right|}\right).
\end{equation}

In the usual case of a rock about a vertical or horizontal axis, $\frac{\left|\vec{q}\times\vec{r}\right|} {\left|\vec{r}\right|}$ is equal to $q_x$ or $q_y$ respectively.

For a set of measurements $\{D_{ij}\}$ at a given pixel $i$, we want to know the best estimate of $I_i$, i.e. what is the conditional probability $P\left(I_i|\{D_{ij}\},\h\right)$ where $\h$ represents our model and background knowledge, and includes things like $\Gamma$ and the measurement errors. Items to the right of the $|$ sign indicate known information which may affect the probability of $I_i$. Including $\h$ indicates there may be additional relevant parameters which can be explicitly taken into account if need be. We will take the approach of introducing parameters when it becomes apparent they are necessary, which is intended to minimise the amount of computation needed.
\subsection{Single pixel}
Bayes' theorem gives, for a single frame and pixel:

\begin{equation}\label{Bayes}
    P\left(I_i|\{D_{ij}\},\h\right)= \frac{P\left(\{D_{ij}\}|I_i,\h\right)P\left(I_i|\h\right)}{P\left(\{D_{ij}\}|\h\right)}
\end{equation}
$P\left(\{D_{ij}\}|I_i,\h\right)$ is known as the likelihood,
$P\left(I_i|\h\right)$ is the prior, representing our starting knowledge in the form of a probability distribution. $P\left(\{D_{ij}\}|\h\right)$ can usually  be treated as a normalizing constant. This does not depend on $I_i$, and ensures $\int P\left(I_i|\{D_{ij}\},\h\right)dI_i=1$.

The data will have a probability distribution which is a convolution of two Poisson distributions from the foreground and background. If the number of counts is $\gtrsim 10$, this will approximate closely to a Gaussian of mean $f_{ij} I_i$ and variance $\sigma_{ij}^2$ equal to the sum of foreground and background variances.

For a single frame $j$, the likelihood is given by
\begin{equation}\label{likelihood}
   P\left(D_{ij}|I_i,\h\right) = (2\pi\sigma_{ij}^2)^{-1/2}\exp\left[-\frac{(D_{ij}-f_{ij} . I_i)^2}{2\sigma_{ij}^2}\right]
\end{equation}

Let's assume the prior $P\left(I_i|\h\right)$ is a Gaussian with mean $\mu$ and variance $\xi^2$, i.e.
\begin{equation}\label{eq:prior}
    P\left(I_i|\h\right) = (2\pi\xi^2)^{-1/2}\exp\left[-\frac{(I_i-\mu)^2}{2\xi^2}\right].
\end{equation}
The justification for this will be discussed below, but note that $\xi$ has different units from the sample error $\sigma_{ij}$ and represents the prior uncertainty on the total integrated intensity over the entire rock, not the measured intensity in any frame.

It can be shown that because each measurement is conditionally independent, i.e. given $I_i$, any measurement gives no further information about the next, then likelihoods can be multiplied\cite{Dawid79}:
\begin{gather}
   P\left(\{D_{ij}\}|I_i,\h\right) = \prod_j P\left(D_{ij}|I_i,\h\right)\nonumber \\
    = \textrm{const.}\times\prod_j \exp\left[-\frac{(D_{ij}-f_{ij} I_i)^2}{2\sigma_{ij}^2}\right] \nonumber \\
    =\textrm{const.}\times\exp\left[-\frac{1}{2}\sum_j\left(\frac{D_{ij}-f_{ij} I_i}{\sigma_{ij}}\right)^2\right].\label{eq:allframes}
\end{gather}

Substituting \eqref{eq:prior} and \eqref{eq:allframes} into Eq.~\eqref{Bayes}, remembering that the denominator is a normalising constant, gives:
 \begin{multline}\label{eq:posterior}
   P\left(I_i|\{D_{ij}\},\h\right) =  \textrm{const.}\times\\ \exp\left\{-\frac{1}{2}\left[\left(\frac{I_i-\mu}{\xi}\right)^2+\sum_j\left(\frac{D_{ij}-f_{ij} I_i}{\sigma_{ij}}\right)^2\right]\right\}
 \end{multline}
To find the optimum value of $I_i$, we maximise \eqref{eq:posterior}, or equivalently its logarithm:

\begin{multline}\label{eq:logP}
-\frac{1}{2}\frac{\partial}{\partial I_i}\left[\left(\frac{I_i-\mu}{\xi}\right)^2+\sum_j\left(\frac{D_{ij}-f_{ij} I_i}{\sigma_{ij}}\right)^2\right] \\
= \frac{I_i-\mu}{\xi^2}+\sum_j \left[ \frac{-f_{ij}(D_{ij}-f_{ij} I_i)}{\sigma_{ij}^2}\right] = 0
\end{multline}
Collecting terms in $I_i$:
\begin{equation}
    I_i\left(\frac{1}{\xi^2}+\sum_j\frac{(f_{ij})^2}{\sigma_{ij}^2}\right) = \frac{\mu}{\xi^2}+\sum_j \frac{f_{ij}D_{ij}}{\sigma_{ij}^2}
\end{equation}

i.e. the value of $I_i$ which maximises $P\left(I_i|\{D_{ij}\},\h\right)$ is
\begin{equation}\label{bestfit}
    I_i= \frac{\mu/\xi^2+\sum_j f_{ij}D_{ij}/\sigma_{ij}^2}{1/\xi^2+\sum_j{f_{ij}^2}/{\sigma_{ij}^2}}
\end{equation}

This can be written as
\begin{equation}\label{eq:weights}
    I_i = I_0 + \sum_j w_{ij} D_{ij}
\end{equation}
where $I_0$ is proportional to the prior mean and uniform across all pixels, and the weights $w_{ij}$ are given by
\begin{equation}\label{eq:weights2}
    w_{ij} = \frac{f_{ij}/\sigma_{ij}^2}{1/\xi^2+\sum_j{f_{ij}^2}/{\sigma_{ij}^2}} .
\end{equation}

The error on $I_i$ can be found by assuming that the posterior probability is Gaussian, which is true in this case as it is formed from the product of Gaussian functions. For a Gaussian function $f(x)$ with mean $x_0$, the variance is given by $f(x_0)/f''(x_0)$, i.e.

\begin{equation}
  \sigma_{\textrm{total}}^2 = \left(\frac{1}{\xi^2}+\sum_j \frac{f_{ij}^2}{\sigma_{ij}^2}\right)^{-1}
\end{equation}

So the best estimate of $I_i$, including error is given by:
\begin{equation}\label{eq:bestfiterror}
    I_i = \frac{\mu/\xi^2+\sum_j f_{ij}D_{ij}/\sigma_{ij}^2}{1/\xi^2+\sum_j{f_{ij}^2}/{\sigma_{ij}^2}}\pm \left(\frac{1}{\xi^2}+\sum_j \frac{f_{ij}^2}{\sigma_{ij}^2}\right)^{-1/2}
\end{equation}

\subsection{Multiple pixels}
When dealing with the entire detector, we can do the same thing for each pixel. The values of $f_{ij}$ will be different for each position, and depend on the geometry of the setup, the neutron wavelength, and the value of $\Gamma$.  The GRASP data analysis package \cite{GRASP}, which was used to display and process the data (i.e. subtract backgrounds, normalise by total monitor counts etc.), provides information about the values of $\vec{q}$ at each pixel position, and information about the Ewald sphere.  This can be used to calculate $q_z$ for any pixel, and hence $f_{ij}$, given $\Gamma,\delta_\omega$, and $\delta_\phi$, where we introduce the latter to represent any misalignment of the field with the coordinate axes defined in the experimental setup. These are included by subtraction from the diffractometer angles reported in the measurement file.

Usually we wish to know the optimum values of $\{I_i\},\Gamma,\delta_\omega$, and $\delta_\phi$, i.e. those which maximise the posterior probability. To find these we make the rather strong simplifying assumption that all pixels are independent, implying their individual probabilities can be multiplied to form a joint posterior probability for all pixels. First, the optimum integrated intensity $\{I_i\}$ for each pixel is calculated analytically as a function of $\Gamma,\delta_\omega,\delta_\phi$ using Eq.~\eqref{eq:bestfiterror}, and the probability for each pixel by Eq.~\eqref{eq:posterior}. It is quite straightforward to then numerically maximise the (log) joint posterior probability with respect to $\Gamma,\delta_\omega,\delta_\phi$, using priors which are broad Gaussians encompassing all likely values for all three.

When diffraction spots are visible in the raw data, which is usually the case, it is useful to mask the regions without diffracted intensity and exclude them from the probability calculation. When the optimimum values of $\Gamma,\delta_\omega$, and $\delta_\phi$ have been determined, these can be used to calculate the entire diffraction pattern.

The independent pixel assumption does not take account of the finite in-plane instrumental resolution, which dominates the spot shape. The present method therefore preserves in-plane correlations. This should not affect the values of, for example, total integrated intensities over a whole diffraction spot, but may mean that errors are underestimated.

The misalignment $\delta_\omega,\delta_\phi$ depends on the experimental setup, not the sample, except in cases where pinning to twin planes is extremely strong. It can therefore be determined using a reference sample which gives a strong signal, such as niobium, and regarded as fixed for subsequent measurements.

The errors $\{\sigma_{ij}\}$ should be averaged over all frames for each pixel. This reduces noise and avoids unwanted correlations between weights and frame number.

\subsection{Choice of prior}\label{section:prior}

In general we do not want our prior to significantly influence the result in regions where the data is informative, i.e. where the Ewald sphere has crossed the reciprocal lattice plane. This requirement means that the prior variance should be significantly larger than that of data, and preferably should have a significant probability at the `correct' intensity. We also wish to avoid introducing any extra structure, so it must be applied uniformly across the detector.  This can be accomplished while still leaving some freedom in areas where the data is not very informative. There is more than one approach we can take, each with different advantages; the choice is largely determined by how we would like to present the result, ideally in a pleasing, yet accurate manner.

Beyond its 2D nature, for a FLL diffraction pattern the prior information we have is this: most pixels to have zero intensity, with a finite number of spots at $q$-vectors with a simple symmetry and a positive intensity corresponding to the highest count value found in any individual frame. To reflect this accurately would require a complicated prior which assumes rather too much about the form of the diffraction pattern.  Therefore for reasons of transparency and ease of calculation we will use the Gaussian prior shown in Eq.~\eqref{eq:prior}, with mean $\mu=0$, and variance $\xi^2$ chosen large enough to encompass the range of intensities expected. So long as the prior is much broader than the likelihood function of the data, relevant data will quickly dominate the posterior probability density function. Setting $\mu=0$ can lead to negative as well as positive values, but it has the advantage that in the absence of relevant information it gives zero average intensity.

One option to determine $\xi$ is to use a value equal to the maximum intensity recorded in any pixel (excluding the direct beam), normalized by $\pi\Gamma/2$ to give an integrated intensity. This is a fairly good representation of our true state of knowledge, as we do not know if there are any spots in the unmeasured regions. The main problem with this prior is that it still produces rather large amounts of unsightly noise in these areas. This can be dealt with by masking these regions, as described below. Alternatively the prior variance can be reduced until the noise is at an acceptable level, though care should be taken that the measured regions are not significantly affected.

Another possibility is to use a separate prior variance for each pixel, determined in the same way but using (a few times) the maximum number of counts in the individual pixels over the whole range of rocking angles.  This has the effect of reducing the noise and giving a nicer looking result, but it will also tend to suppress weak signals, or those where $f_{ij}$ is small for all angles but there is some evidence of a signal.

Masking of unmeasured regions can be carried out by comparing the prior and posterior probabilities. In regions of the detector in which the Bragg criterion is never reached, the data collected carries little information about the in-plane intensity.  In this case the posterior variance will be close to that of the prior. These areas can be found by taking the ratio of the prior to posterior variances and masking pixels where this ratio is above a certain threshold. Alternatively, the uncertainty on the final result could be encoded in a color mapped display of the results, for example as a color saturation value.

One can also consider a prior based on the peak intensity, rather than the integrated intensity (this is equivalent to normalizing the scaling factors $f_{ij}$ such that $f(0)=1$, but keeping everything else the same). In practice this is very much less useful, due to strong correlations between the peak width and height, so the integrated intensity is a better choice.

As all of these methods are applied identically to each pixel, there is no chance of introducing any additional structure beyond that indicated by the data.  So long as the priors are not too strong, which is satisfied by the methods described, the data will dominate in the `measured' regions and the effect of the prior is confined to the places where the inferred intensity is weakly constrained.

\section{Results}
\begin{figure*}
  \subfloat[]{\includegraphics[width=0.9\columnwidth]{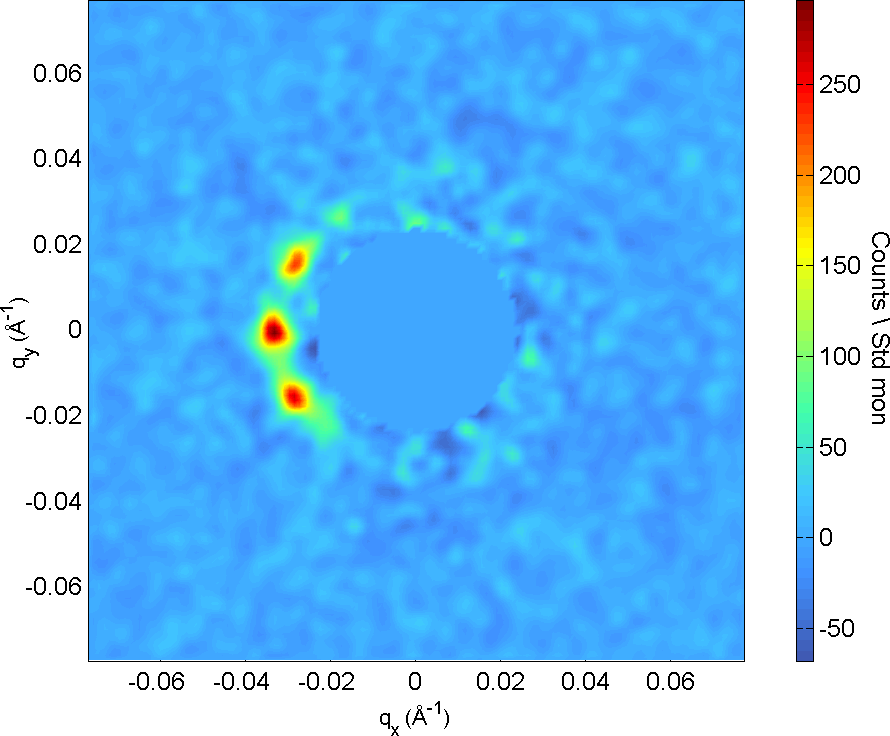}\label{subfig:singleframe}} \qquad \subfloat[]{\includegraphics[width=0.9\columnwidth]{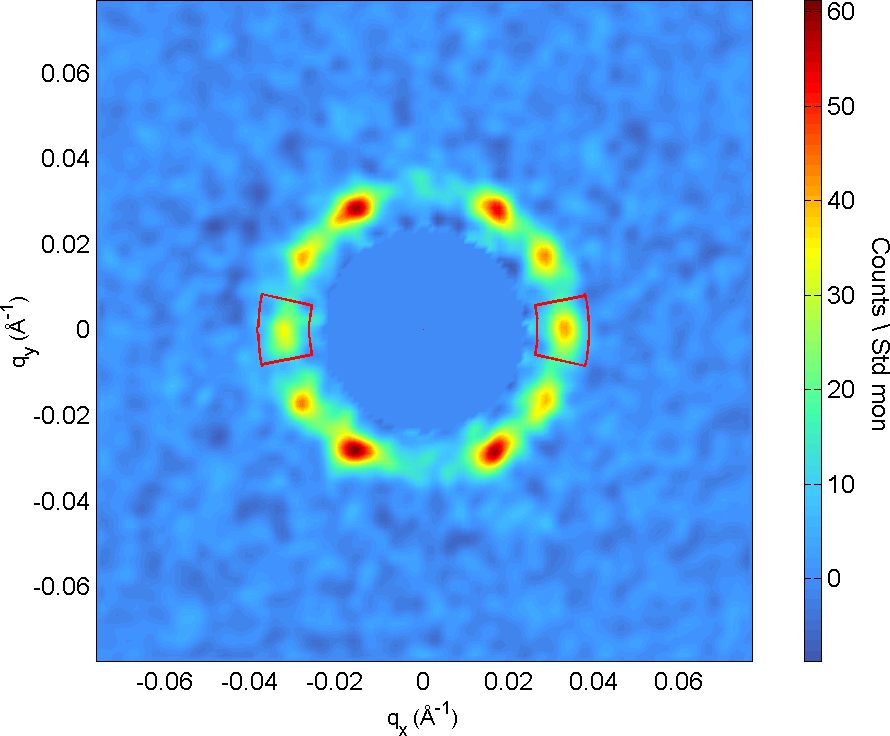}\label{subfig:fullsum}}\\ \subfloat[]{\includegraphics[width=0.9\columnwidth]{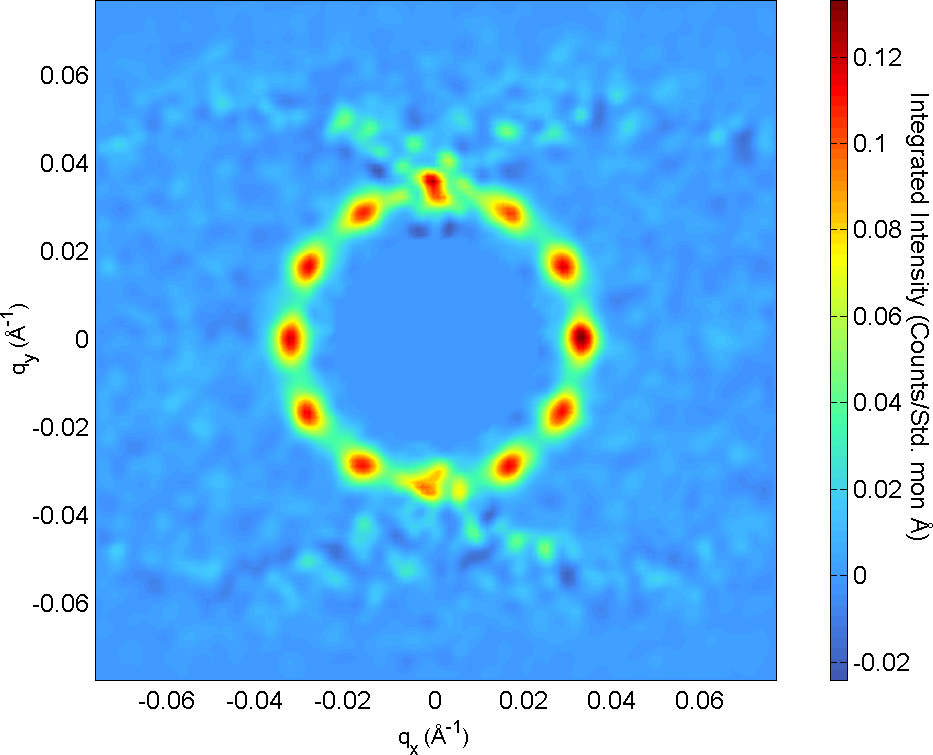}\label{subfig:pixelprior}}   \qquad \subfloat[]{\includegraphics[width=0.9\columnwidth]{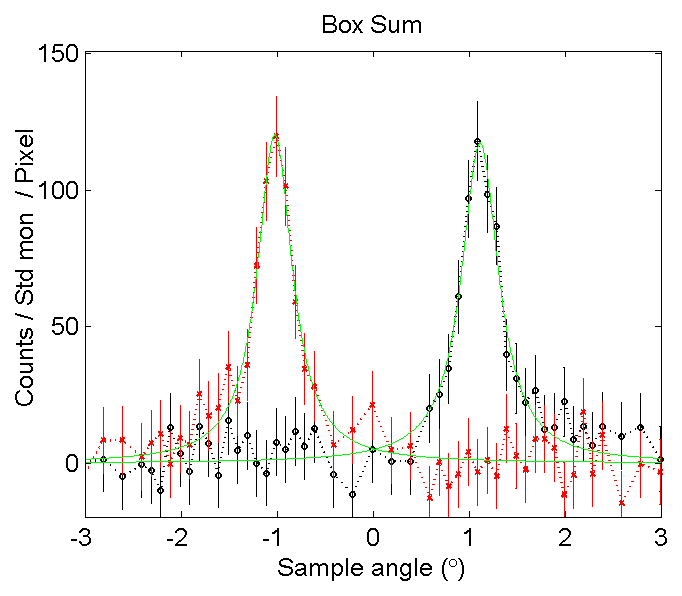}\label{subfig:rockingcurve}}
  \caption{(Color online) \protect\subref{subfig:singleframe} Data taken at a single angle $\omega$. Only certain spots fulfil the Bragg condition to within the experimental resolution. \protect\subref{subfig:fullsum} Sum of data taken over a range of angles encompassing all first order Bragg reflections attainable by rotation about a vertical axis. \protect\subref{subfig:pixelprior} Result of Bayesian weighted method, using individual pixel priors, as described in section \ref{section:prior}. \protect\subref{subfig:rockingcurve} Sum of intensity inside boxes shown in \protect\subref{subfig:fullsum} plotted as a function of rocking angle. The width of the peaks is a measurement of the longitudinal coherence of the flux lattice, and the total integrated intensity under the peaks is proportional to the square of $B(\vec{q})$. The data in both \protect\subref{subfig:singleframe} and \protect\subref{subfig:fullsum} have backgrounds taken in the normal state subtracted from them. A small amount of smoothing (convolution with a $3\times3$ pixel Gaussian) has been applied to \protect\subref{subfig:singleframe}-\protect\subref{subfig:pixelprior} after all other processing. The direct beam in the centre of the detector is masked.}\label{fig:BFAPdata}
\end{figure*}

Figure \ref{fig:BFAPdata} shows diffraction patterns from BaFeAs$_{(1-x)}$P$_x$ ($x=0.3$) at 5$\:$T, obtained by rocking about a vertical axis \cite{MorisakiIshii14}.  Note in \ref{fig:BFAPdata}\subref{subfig:fullsum} the difference in intensity of the on-axis and off-axis spots, and the absence of spots at the top and bottom.  Figure \ref{fig:BFAPdata}\subref{subfig:rockingcurve} shows the summed intensities in the marked sector boxes as a function of angle $\omega$.  Off-axis spots will have larger FWHM, by a factor of $1/|\textrm{cos}(\alpha)|$, where $\alpha$ is defined above. This means that they will accumulate more intensity in a summed rock.

\begin{figure}
  \subfloat[]{\includegraphics[width=0.7\columnwidth]{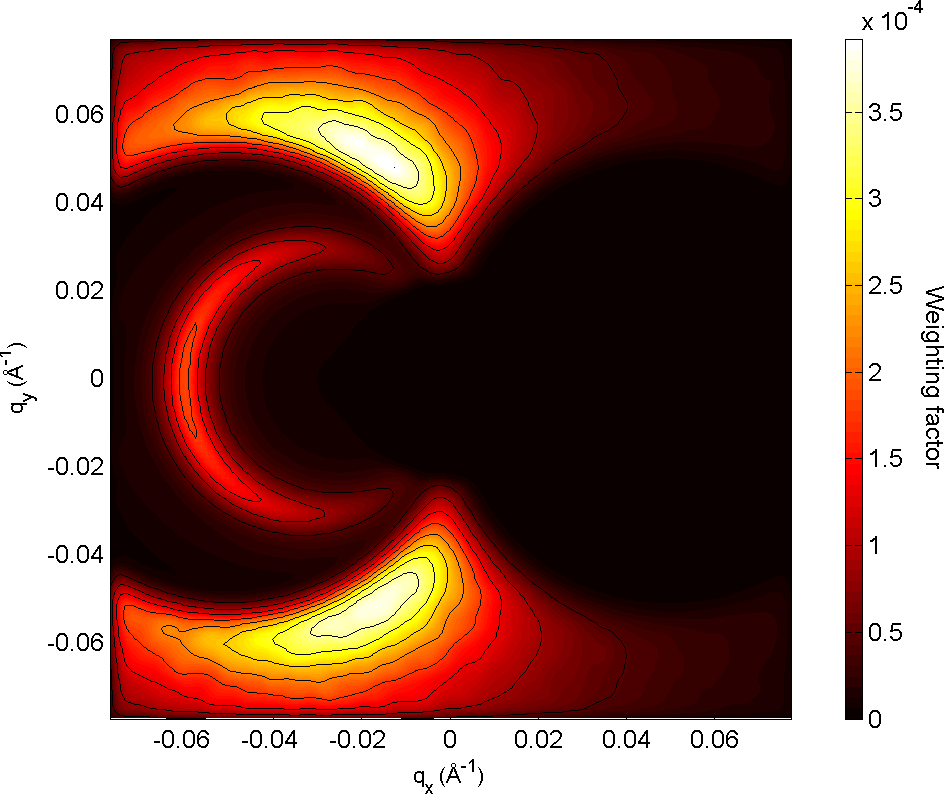}\label{subfig:weightsuniform}}\\
  \subfloat[]{\includegraphics[width=0.7\columnwidth]{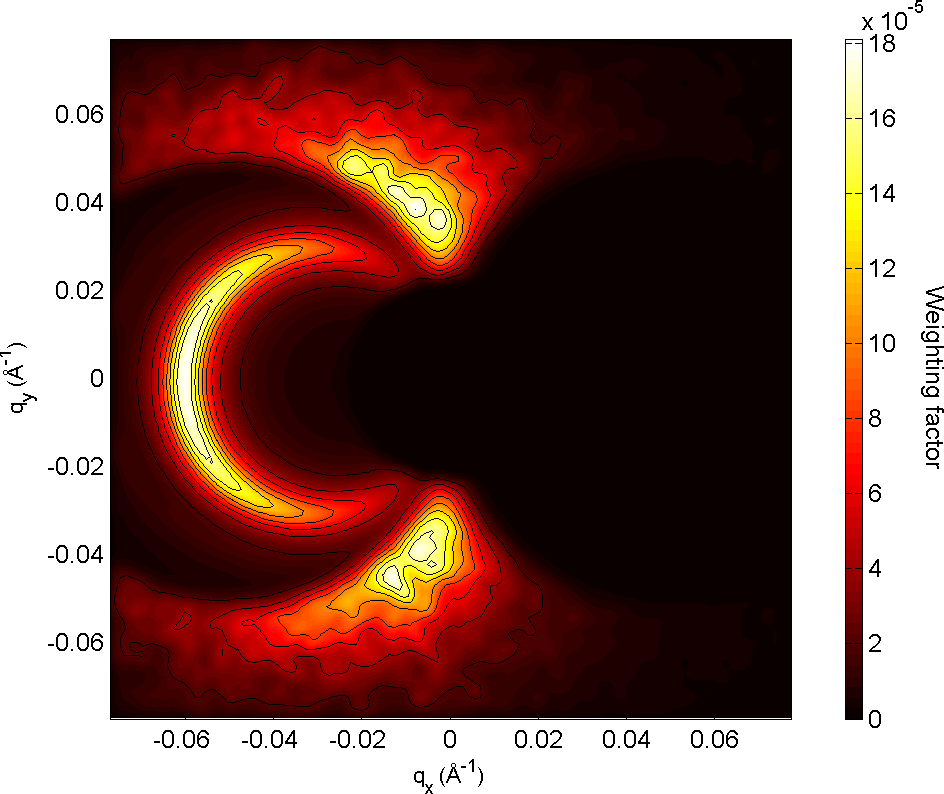}\label{subfig:weightspixel}}  \caption{(Color online) Weighting factors $w_{ij}$ for a single frame using \protect\subref{subfig:weightsuniform} uniform prior, and \protect\subref{subfig:weightspixel} individual pixel priors, as described in section \ref{section:prior}.  Note the crescent shape where the Ewald sphere cuts the lattice place, but also the maxima at the top and bottom, which are the result of the denominator in equation \ref{eq:weights2}. In \protect\subref{subfig:weightspixel} these are less prominent as the prior is suppressed by the low counts in these regions, though more noise is evident. $3\times3$ pixel Gaussian smoothing has been applied to both figures.}\label{weights}
\end{figure}

Figure \ref{weights} shows a typical map of the weighting coefficients ${w_{ij}}$ for a frame taken at a single angle from a rock about the $y$-axis, calculated using Eq.~\eqref{eq:weights2}.  As expected, there is a maximum corresponding to the position where the Ewald sphere intersects the reciprocal lattice. More surprisingly, there is also a region around the rocking axis, where a large weighting factor appears.  This is due to a small value of the denominator in \eqref{eq:weights2}, when all values of $f_{ij}$ are $\ll 1$ but the prior does not quite dominate. If the rocking curve width is broad, there may be relevant data present from the tails of the peak, and a diffraction spot can be recovered which was not actually measured!  Another slightly counterintuitive effect that can occur, particularly with narrow rocking widths or rocks in more than one direction, is that maxima in the weighting factor of one frame produces minima in the others, again via the denominator of \eqref{eq:weights2}.

\begin{figure*}
  \subfloat[]{\includegraphics[width=0.9\columnwidth]{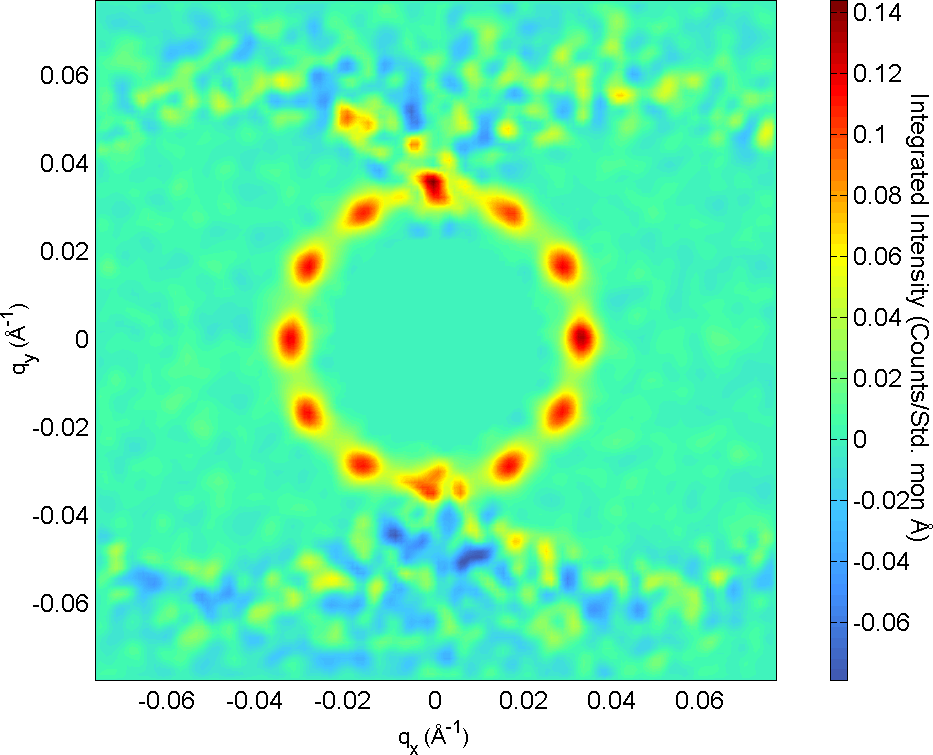}\label{subfig:bayesed_sum_uniform}}\qquad
  \subfloat[]{\includegraphics[width=0.9\columnwidth]{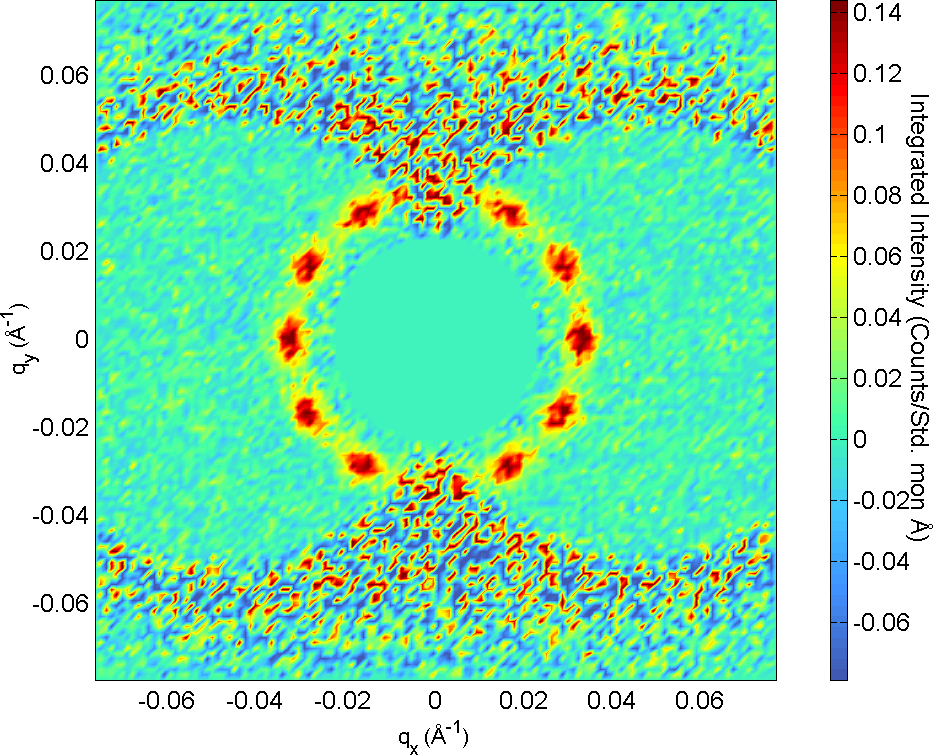}\label{subfig:bayesed_sum_nosmooth}}\\
  \subfloat[]{\includegraphics[width=0.9\columnwidth]{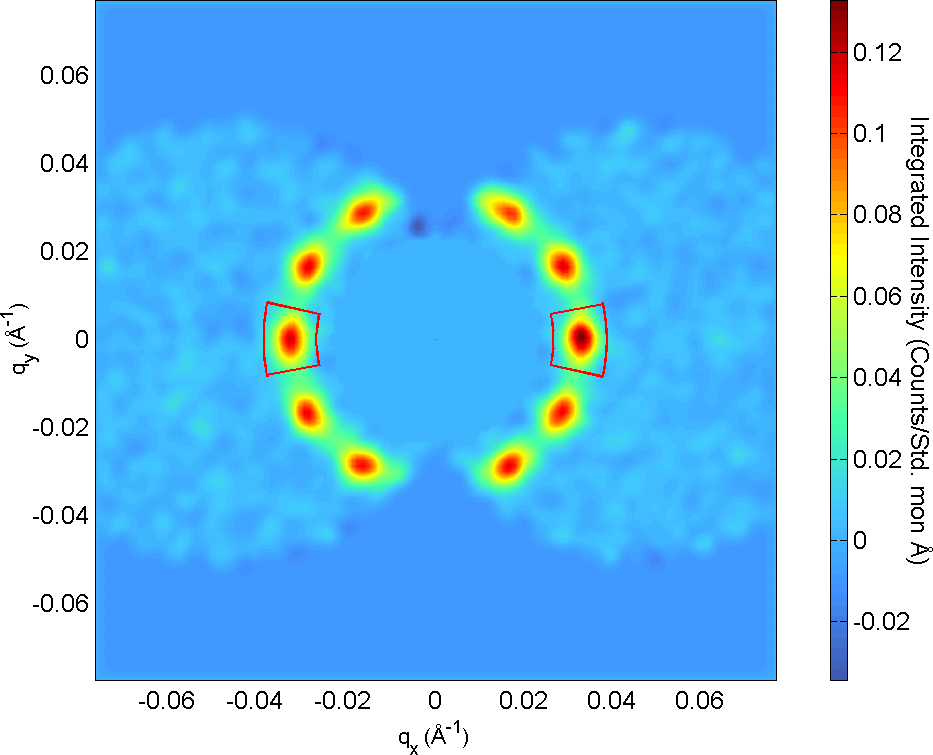}\label{subfig:bayesed_sum_masked}}\qquad
  \subfloat[]{\includegraphics[width=0.9\columnwidth]{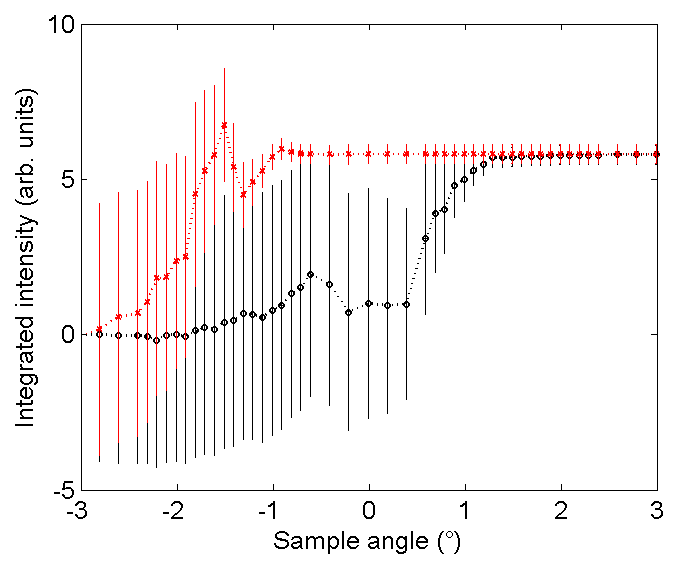}\label{subfig:cumul_box_sum}}

  \caption{(Color online) \protect\subref{subfig:bayesed_sum_uniform} Result of Bayesian weighted sum of data from Fig.~\ref{fig:BFAPdata}, with prior mean $\mu$ set to zero, and uniform s.d. $\xi$ determined by the maximum spot intensity. Note that spots are recovered at the top and bottom, even though they never reach the Bragg criterion. Gaussian smoothing has been applied to the weighted sum with a $3\times3$  pixel resolution. \protect\subref{subfig:bayesed_sum_nosmooth} The same as  \protect\subref{subfig:bayesed_sum_uniform} without smoothing. \protect\subref{subfig:bayesed_sum_masked} Areas with $\sigma > 0.2\xi$ are masked, where $\sigma$ is the error on the posterior intensity. \protect\subref{subfig:cumul_box_sum} sum of intensity inside boxes in \protect\subref{subfig:bayesed_sum_masked} as a function of angle. These converge on the final result at different positions, depending on the angle at which the peak appears. Data taken at angles after passing through the peak of the rocking curve at the Bragg condition make very little difference to the final result.}\label{BFAP 5T}
\end{figure*}

Figure \ref{BFAP 5T} shows the result using original data from \ref{fig:BFAPdata}, with and without smoothing (applied after all processing) and masking.  In Fig.~\ref{BFAP 5T}\subref{subfig:cumul_box_sum} the cumulative estimate for the integrated intensity of two on-axis diffraction spots is shown, along with error bars corresponding to $\pm 1\sigma$. As the data arrives, the estimated intensity of each spot converges, within errors, to a single value. In this example only one, vertical, rocking axis has been used, partly to illustrate the results with incomplete reciprocal space coverage. This data can be combined with that from a rock about a horizontal axis to give a more complete set of information about all spots with useful intensity over the entire detector area.

It is evident from inspection that this method gives a significant improvement in the signal to noise ratio. However it is hard to give this an exact value, as it relates to the fraction of the rocking curve which contains relevant data, which varies across the detector. One can make an estimate for the data shown here of about a factor of two, thanks to the narrow rocking curve width.

An implementation of this method using Matlab\textsuperscript{\textregistered}\ is available from the author.

\section{Discussion}
\subsection{Simplifying assumptions}

The method shown above is deliberately as simple as possible, so as to be easily understood and reproduced as required.  It is subject to the following approximations and limitations:
\begin{itemize}
  \item The diffraction pattern is treated as strictly 2D, plus a single perpendicular coherence length. This could easily be extended to different coherence lengths along different axes, but it will fail to correctly account for any out of plane scattering beyond this.
  \item Probability distributions for data are initially treated as Gaussian - in reality the foreground and background should be treated as two separate Poissons, possibly with different scaling factors.
  \item No account of finite in-plane resolution is taken.  This could be incorporated into further modelling of the diffraction pattern, or treated as a deconvolution problem. The presence of correlations between pixels means that the calculated errors on `box sums', summed intensity over several pixels, may be incorrect, however the optimum values should reflect the true integrated intensity (again assuming everything is Gaussian).
  \item The maximum posterior probabilities for the rocking width and misalignment values are used.  In principle with additional computational effort one could marginalize these out as nuisance parameters i.e.\ integrate over all possible values weighted by the posterior probability.
\end{itemize}

A way in which the presented method can fail to improve on the usual unweighted sum is if there is a very broad rocking curve, much larger than the range of angles measured. In this case all frames will contain relevant information, so there will not be a large improvement in the signal to noise ratio. The misalignment and rocking curve width will not be strongly constrained by the data, so unless there is information about this from other sources, the weighting factors may not be accurate, even for the best fit solution. This is actually somewhat reassuring, as the Bayesian method will not make a bad sample into a good one!

\subsection{Philosophical issues}
Many objections to Bayesian methods relate to so-called `subjective' priors. The prior is, however, unavoidable, as any other approach is equivalent to an implicit, usually uniform prior. In the limit of very large amounts of high quality data, the choice of prior will be largely irrelevant, however in scattering experiments we are often limited by statistical noise, and we wish to extract the maximum information from data collected in a limited region of reciprocal space with finite resolution. In this case choice of a suitable prior can be extremely useful.

In the present case we can divide the prior into two components. The first is the prior as understood in the usual sense of a prior probability distribution for the parameters of interest, i.e. $P\left(\{I_i\},\Gamma,\delta_{\theta},\delta_{\phi}|\h\right)$, which can be separated as $P(\{I_i\}|\h)P(\Gamma|\h)P(\delta_{\theta}|\h)P(\delta_{\phi}|\h)$ if all are independent.  In general, so long as no zero values are included, and the sharpness or curvature (i.e. information content) of the prior distribution is much smaller than that of the likelihood, this part of the prior will make very little difference to the final result after accounting for the data. On the other hand, using a broad but informative prior encompassing a physically reasonable range of parameters can help the solution to converge numerically in a way that may not happen if a uniform prior is chosen and the data is not good enough in quality or quantity. Previous measurements carried out with the same or different techniques can also be used to produce an informative prior. This can be very useful, as often different measurement methods (X-ray vs neutron scattering for example) can have different sensitivities or resolution in different parts of parameter space.

The second component to the prior is more subtle, it essentially comprises the choice of model (or set of models) to be considered, and contains information about symmetries of the system. This is contained within the background information $\h$.

The combination of the two types of prior determine the parameter space of interest; the larger the parameter space, the more data required to come to a conclusion. This means in particular that prior information about symmetry can have a huge effect. In the current case the reduction from 3-D to 2-D reciprocal space means a large increase in the signal to noise ratio, as one can average over the irrelevant dimensions.

An interesting question remains as to whether or not the entire background information $\h$ can be described in this way. That is to say, what information is necessary to fully define a statement of probability as a state of knowledge, once a record of all experimental results has been taken into account? Is the choice of model space and symmetry enough, or is some other information required?  This is a philosophical, or perhaps mathematical point, but raises an important issue when considering how to represent actual probabilities, which in practice are always conditional, in an unambiguous way.

\section{Conclusions}

The results shown on flux lattices demonstrate that Bayesian techniques can provide very large improvements in the quality of data analysis. The resulting scattering patterns more accurately represent the integrated intensities than simple sums of the data. They can include the usual (Lorentz) corrections in a natural way, by working in a model space suited to the problem.  A very simple Gaussian analytical treatment is more than adequate. This has the advantage of transparency as well as speed of implementation. The maximum posterior probability solution was chosen, with rocking curve width and field misalignments with the coordinate axes left as free parameters.

The case of the FLL may be particularly suited to this problem, due to its 2-D nature, but it is by no means the only such system, another example being Skyrmion lattices \cite{Muehlbauer09}. Often there is also a lot of accumulated data about instrument characteristics, or accumulated experimental data from previous experiments.  These can all be naturally taken into account using Bayesian methods, and there are many examples of such work over the years in a wide variety of fields \cite{Sivia92,Klementev01,Dose03,Hogg10,vonToussaint11}.  Due to the relatively complex nature of the analysis, which often gives similar if not identical results to more traditional methods (though the exceptions are of significant interest), this has mostly been carried out by motivated individuals with a model hard-coded into the analysis.  However the methods themselves are completely general and model independent.  As computers get cheaper and more powerful with respect to the cost of gathering data, it is starting to make sense to provide general tools for model-independent Bayesian data analysis.  By separating the modeling from the calculations and using standardized methods for data reduction it would no longer be necessary to reinvent the wheel for each problem. A dataflow programming language, such as Labview/G would be the most natural way of encapsulating this process, particularly as many of the calculations are inherently parallel. This would require an agreed data format for conditional probability distributions, meeting the requirements detailed above to include all necessary background information. This would particularly suit large facility-based techniques such as neutron and synchrotron scattering, which have institutional computing support and a well characterized data archival procedure.  We have shown how this method is highly advantageous for one particular system, and look forward to its widespread adoption.

\begin{acknowledgments}
Many thanks to Ted (E.M.) Forgan for the original suggestion of a weighted rocking sum and many thoughtful discussions, Charles Dewhurst for letting me butcher his GRASP software, Hazuki Kawano-Furukawa for generous use of her BFAP SANS data, Jonny White, Sebastian M\"{u}hlbauer and Elizabeth Blackburn for useful discussions and comments.
\end{acknowledgments}
\bibliography{bayes_refs}

\end{document}